\documentclass[journal,twoside,web]{ieeecolor}
\usepackage{jsen}
\usepackage{cite}
\usepackage{graphicx,dblfloatfix}
\usepackage{amsmath,amssymb,amsfonts}
\usepackage{epstopdf}
\usepackage{latexsym}
\usepackage{algorithmic}
\usepackage{graphicx}
\usepackage{textcomp}
\usepackage{wrapfig}
\usepackage{fixltx2e}
\usepackage{multirow}
\usepackage{lineno}
\usepackage{url}
\def\BibTeX{{\rm B\kern-.05em{\sc i\kern-.025em b}\kern-.08em
    T\kern-.1667em\lower.7ex\hbox{E}\kern-.125emX}}
\begin{document}
\title{A Fully Integrated Sensor-Brain-Machine Interface System for Restoring Somatosensation}
\author{Xilin~Liu,~\IEEEmembership{Member,~IEEE,}
        Hongjie~Zhu,~\IEEEmembership{Member,~IEEE,}
        Tian~Qiu,
        Srihari~Y. Sritharan,
        Dengteng~Ge,
        Shu~Yang,
        Milin~Zhang,~\IEEEmembership{Senior Member,~IEEE,}
        Andrew~G. Richardson,~\IEEEmembership{Senior Member,~IEEE,}
        Timothy~H. Lucas,~\IEEEmembership{Member,~IEEE,}
        Nader~Engheta,~\IEEEmembership{Life Fellow,~IEEE,}
        and~Jan~Van der Spiegel,~\IEEEmembership{Life Fellow,~IEEE}
\thanks{
This work was supported by National Science Foundation grant CBET-1404041.}
\thanks{Xilin~Liu, Hongjie~Zhu, Tian~Qiu, Nader~Engheta, and Jan~Van der Spiegel are with the Department of Electrical and Systems Engineering, University of Pennsylvania, Philadelphia, PA, 19104 USA. }
\thanks{Milin~Zhang is with the Department of Electronic Engineering, Tsinghua University, Beijing, China, 100084.}
\thanks{Srihari~Y. Sritharan, Andrew~G. Richardson, and Timothy~H. Lucas are with the Department of Neurosurgery, University of Pennsylvania, Philadelphia, PA, 19104 USA.}
\thanks{Dengteng~Ge is with the Institute of Functional Materials, Donghua University, Shanghai, China, 200336.}
\thanks{Shu~Yang is with the Department of Materials Science and Engineering, University of Pennsylvania, Philadelphia, PA, 19104 USA. }
}

\IEEEtitleabstractindextext{\begin{wrapfigure}[15]{r}{3.5in}%
\includegraphics[width=3.5in]{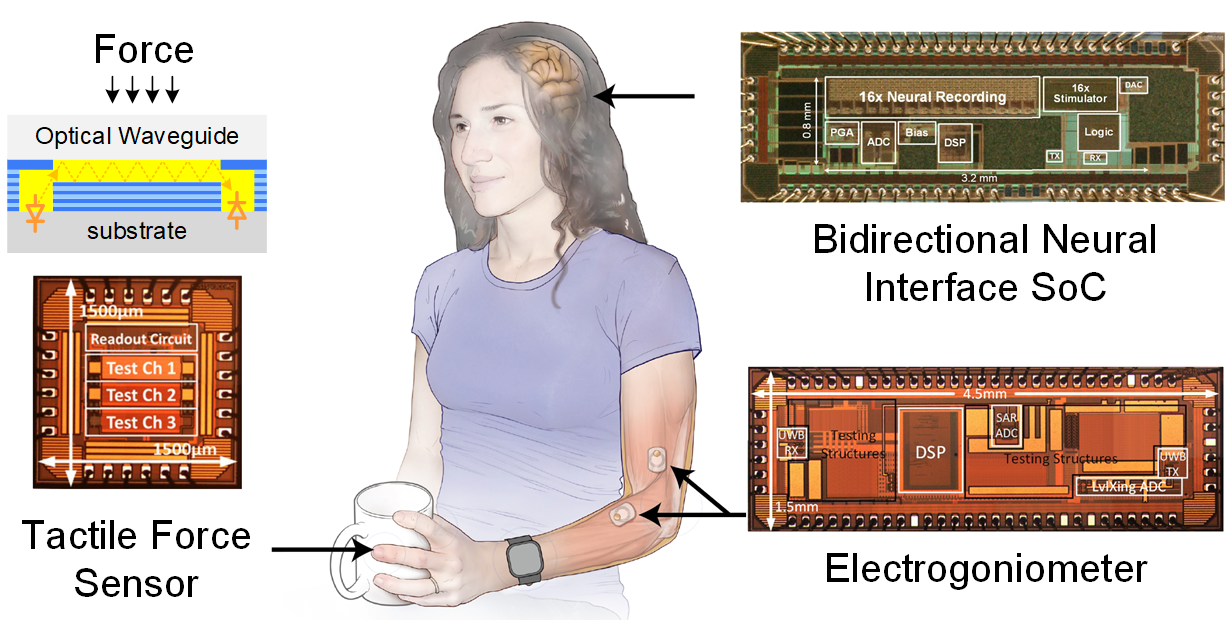}%
\end{wrapfigure}%
\begin{abstract}
Sensory feedback is critical to the performance of neural prostheses that restore movement control after neurological injury. Recent advances in direct neural control of paralyzed arms present new requirements for miniaturized, low-power sensor systems. To address this challenge, we developed a fully-integrated wireless sensor-brain-machine interface (SBMI) system for communicating key somatosensory signals, fingertip forces and limb joint angles, to the brain. The system consists of a tactile force sensor, an electrogoniometer, and a neural interface. The tactile force sensor features a novel optical waveguide on CMOS design for sensing. The electrogoniometer integrates an ultra low-power digital signal processor (DSP) for real-time joint angle measurement. The neural interface enables bidirectional neural stimulation and recording. Innovative designs of sensors and sensing interfaces, analog-to-digital converters (ADC) and ultra wide-band (UWB) wireless transceivers have been developed. The prototypes have been fabricated in 180nm standard CMOS technology and tested on the bench and \emph{in vivo}. The developed system provides a novel solution for providing somatosensory feedback to next-generation neural prostheses.
\end{abstract}

}

\maketitle

\section{Introduction}
\label{sec:introduction}
\IEEEPARstart{S}{omatosensation} --- the sense of touch and posture derived from mechanoreceptors in the skin, muscles, and joints --- is especially important for dexterous control of hand and limb movements \cite{Witney2014, richardson2016effects}. Correspondingly, this sensory feedback is also important for prosthetic systems designed to replace hand and limb function following amputation or paralysis.

Neural prostheses can restore functional movements after injury by establishing a brain-machine interface (BMI). These systems decode movement-related information from neural recordings and transform it into control commands to drive a robotic arm \cite{hochberg2012reach}. In most BMI demonstrations, the sole feedback to the user is the visual correspondence between intended and actual movements. However, this does not yield adequate performance in many real-world tasks involving interaction forces with the environment (e.g. grasping and lifting a cup) \cite{lucas2017Strategies}. Thus, robotic arms can be equipped with sensors that transduce somatosensory stimuli \cite{wettels2008}. The sensor output can then be encoded into the brain through electrical stimulation at a point above the injury along the neural pathway that normally processes this information: peripheral nerves \cite{Daniel2014,raspopovic2014restoring}, subcortical nuclei \cite{sritharan2016Somatosensory,loutit2020}, or somatosensory cortex \cite{flesher2016,armenta2018}. We refer to this as a sensor-brain-machine interface (SBMI).

For paralyzed individuals, the user experience of using robotic limbs is suboptimal \cite{blabe2015}. The ideal rehabilitative strategy is to reanimate the person's own paralyzed limb. Recent work has demonstrated that this is possible using brain-controlled functional electrical stimulation of arm and hand muscles \cite{bouton2016,ajiboye2017}. However, the paralyzing injury also prevents the mechanoreceptor signals within the reanimated limb from reaching the brain. Artificial sensors of somatosensory stimuli are again needed to provide the feedback required for skillful movement. Due to the numerous differences from robotic arms, including the potential for actuator (muscle) fatigue and the restriction of device components such as wires, batteries, and sensors to surface (i.e. skin) layers, a new sensor strategy is required for reanimated arms.

\begin{figure}[!ht]
  \centering
  \includegraphics[width=0.4\textwidth]{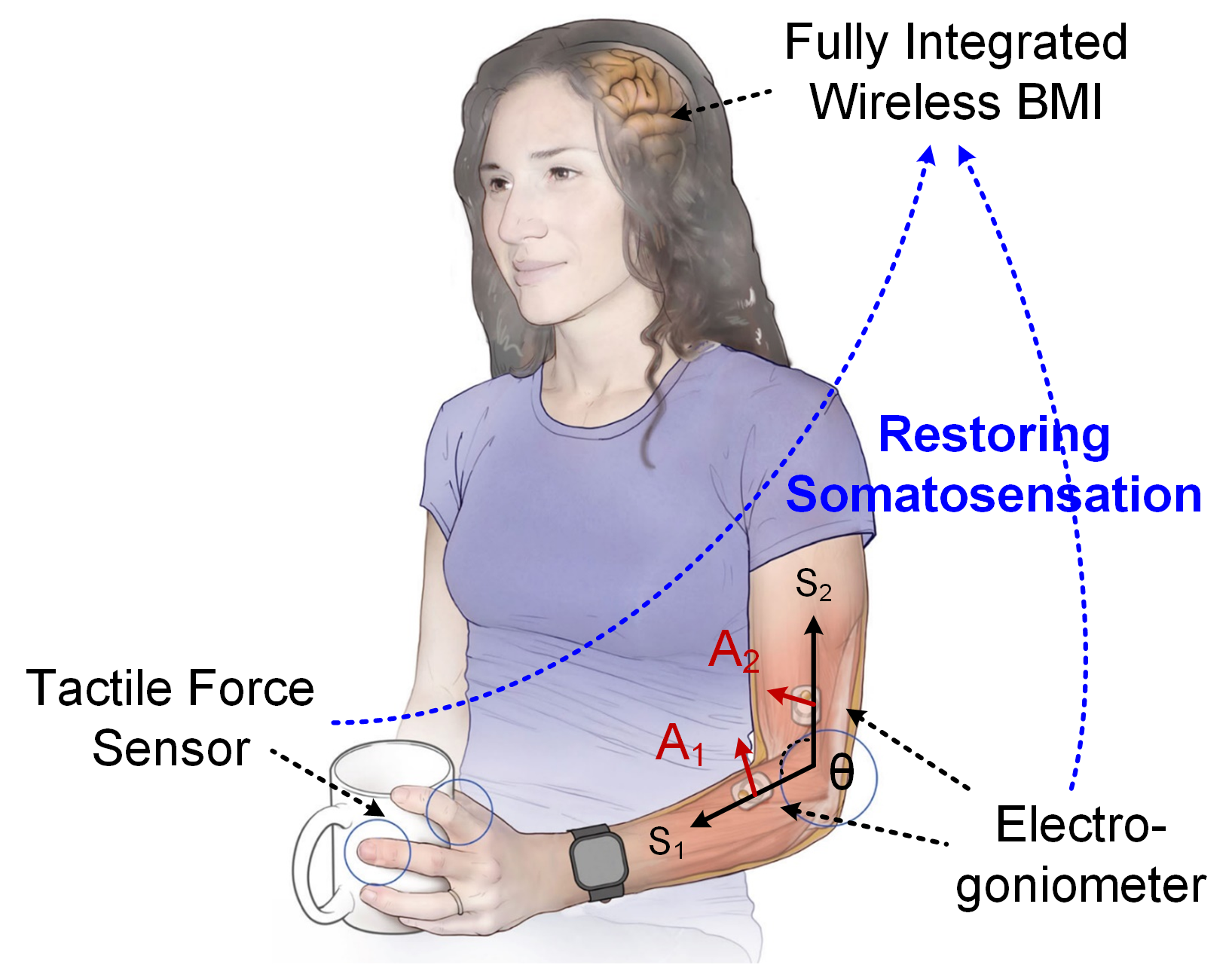}\\
  \caption{Illustration of the proposed wireless sensor-brain-machine-interface (SBMI) system. The system links multiple body-area sensors (e.g. tactile force sensor, electrogoniometer, etc.) and an invasive BMI for continuous somatosensation restoration.}\label{sys_intro}
\end{figure}

In this work, we propose a novel sensor strategy for restoring somatosensation using a SBMI system, as illustrated in Fig. \ref{sys_intro}.
In this new strategy, multiple wireless sensor nodes would be worn on or implanted under the patient's paralyzed limb and have a minimal physical presence, free from the constraints of gloves or wires. Ideally, the sensors would cause minimal additional loads on the muscles and joints and be nearly transparent to the user. A wireless BMI device with an invasive neural interface would provide sensory encoding by continuous electrical stimulation with the stimulus amplitude or frequency modulated by the sensors' outputs. This strategy could provide superior intuitive sensation restoration to paralyzed individuals, but places great demands on the specifications of the sensors and associated electronics, including power consumption and device dimension. To the best of our knowledge, none of the existing commercial medical devices or reported work in literature meet all requirements of this novel sensor strategy. In this work, we fill this important research gap by developing a custom SBMI system with innovative sensors, electronics and system integration. Below, we analyze the design considerations of each key building block.

\subsection{Tactile Force Sensor}
Human skin, and the fingertip in particular, provides high sensitivity to force across a range of frequencies \cite{johansson2009}. Interest in miniature force sensors with high sensitivity has increased in recent years due to their potential applications in electronic skins, touch screens, and medical diagnostics \cite{sekitani2008,chortos2016}. Microelectromechanical systems (MEMS) are one of the fastest-growing technologies in the miniaturized force sensor market \cite{lopez2005}. However, MEMS sensors usually require specialized micro-fabrication processes \cite{zhao2013design,sokhanvar2009mems}, which leads to a high cost. As an alternative, polymer-based optical force sensors have advantages including scalability and insensitivity to electronic noise. Elastomeric polydimethylsiloxane (PDMS) is typically used as a compressible optical cavity or waveguide. Optical pressure sensors consisting of optical fibers and PDMS waveguides have been demonstrated \cite{ramuz2012,missinne2010}. Thanks to advancements in silicon light-emitting devices (Si-LED) \cite{snyman2010}, optical waveguides have become feasible to implement in standard CMOS technology. Using this technology, we develop a wearable optical force sensor with a wireless transmitter.

\subsection{Electrogoniometer}
Proprioception is the sense of the relative position and movement of the body, which is essential for guiding movements. An electrogoniometer is a device used to measure joint angles. It typically employs sensors with relatively high power consumption, such as potentiometers or strain gauges, and thus is not suitable for long-term everyday use \cite{favre2008,palermo2014}. In this work, an electrogoniometer is designed with custom-designed ICs and ultra low-power 3-axis MEMS accelerometers, achieving a very low power consumption and a minimum device dimension. The joint angle is calculated by an on-chip digital signal processor (DSP) using the measurements from the two accelerometers.

\subsection{Bidirectional Neural Interface}
In the past decade, there has been significant progress in the development of bidirectional neural interfaces, which integrate both neural recorder and stimulator \cite{milin2020,jan2016}. Initial studies have demonstrated long-term recording and stimulation in freely behaving animals using off-the-shelf components \cite{mavoor2005,venkatraman2009,zanos2011}. Subsequently, the systems became more integrated to improve performance and enable new applications \cite{Mendrela2016,abdi2016}. A bidirectional BMI was developed in which neural recording and processing subsystems were integrated into a commercial neural stimulator \cite{rouse2011,stanslaski2012}. A system-on-chip (SoC) with 64 recording channels and dual stimulation channels was designed \cite{biederman2015} as was a 32-channel modular bidirectional BMI with an embedded digital signal processor (DSP) for closed-loop operation \cite{cong2014}. Another group developed a battery-powered activity-dependent intracortical microstimulation SoC with on-chip action potential discrimination and spike-triggered stimulation \cite{azin2011}. Follow-up work added an on-chip stimulation artifact rejection feature \cite{limnuson2014}. High channel count designs were developed including a 128-channel fully differential neural recording and stimulation interface \cite{shahrokhi2010} and a 320-channel bidirectional interface chip \cite{shulyzki2015}. Finally, bidirectional neural interfaces have been designed for specific clinical applications, including control of epileptic seizures and movement disorders \cite{chen2014fully,Rhew2014}.

The bidirectional neural interface in the present work builds from prior devices developed by our group. We previously developed a battery-powered, modular system with wireless sensor nodes and a bidirectional neural interface using off-the-shelf components \cite{liu2014PennBMBI,liu2015PennBMBI}. We also developed a fully-integrated bidirectional neural interface SoC with on-chip closed-loop controller \cite{liu2015biocas,liu2016pid}. Here, we extend the latter design to include an ultra-wide band (UWB) wireless link to the sensor nodes and custom on-chip processing to create a low-power, SBMI system that could provide somatosensory feedback for a neural prosthesis.

Phased research progress towards the system described in this paper has been presented previously \cite{zhu2015Design,liu2017Fully}. The contributions of the present paper include the characterization of the optical force sensor and the electrogoniometer, the algorithm and DSP design details underlying the proprioceptive sensing, the methodology and implementation of the system integration, as well as experiments and quantification of the full SBMI system specifications.

\section{System Overview}
The proposed SBMI is a preclinical system with which artificial sensory encoding strategies can be validated in appropriate animal models. In an experiment of finger force encoding, the force sensor detects the fingertip force and sends the readout wirelessly to the neural interface device to deliver modulated stimulation to the brain. Typically, stimulus pulse amplitude or pulse duration would be linearly modulated with the force readout to encode the sense of touch \cite{petrini2019,charkhkar2020}. Similarly, in an experiment of proprioception encoding, the electrogoniometer calculates the joint-angle and sends it wirelessly to the neural interface device for modulated stimulation. Encoding both tactile and proprioceptive stimuli simultaneously could be achieved through separate neural stimulation channels targeting distinct circuits normally responsive to each modality. The neural interface device is worn on the animal's head or back with wired connections to electrodes implanted chronically into target brain regions. Although the recording capabilities of the neural interface SoC are not strictly needed for sensory encoding, they could be used either for movement decoding in a bidirectional neural prosthesis or to monitor stimulus-evoked neural activity for closed-loop encoding strategies \cite{liu2011tnsre}.

\begin{figure}[!ht]
  \centering
  \includegraphics[width=0.5\textwidth]{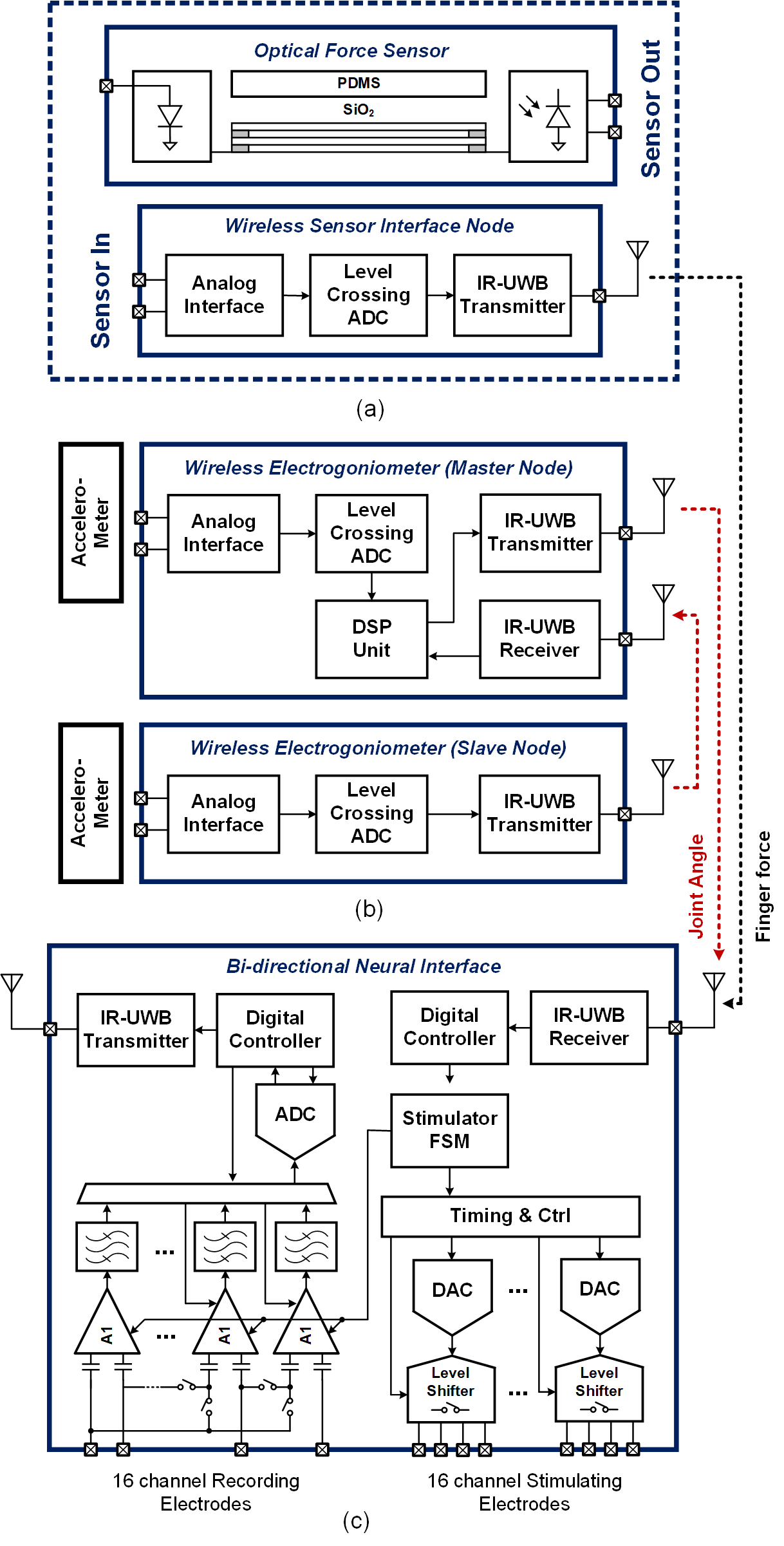}\\
  \caption{Block diagrams of the proposed SBMI system. (a) The wireless optical force sensor, (b) The wireless electrogoniometer sensor pair, and (c) the bi-directional wireless neural interface SoC. Power management units are not shown.}\label{sys_block}
\end{figure}

The block diagrams of the proposed SBMI system are shown in Fig. \ref{sys_block}. The system consists of a wireless tactile force sensor, a wireless electrogoniometer, and a wireless bidirectional neural interface SoC.
The tactile force sensor node consists of a sensor node and a wireless sensor interface IC. The sensor node integrates photodiodes, waveguide channels, and analog readout circuits. The wireless sensor interface IC integrates a programmable analog interface, a low-power asynchronous level-crossing analog-to-digital converter (LxADC), and a UWB transmitter. The two chips are connected on a printed circuit board (PCB).

The electrogoniometer consists of a primary and a secondary node, which are to be attached to two body parts to measure their relative position. Each node has a custom-designed IC and an off-the-shelf 3-axis MEMS accelerometer, which are connected on a PCB. Each node has an analog interface and an ADC to digitize the accelerometer's output. The secondary node integrates a UWB transmitter for transferring the data to the primary node. The primary node integrates a custom-designed DSP, a UWB receiver and a UWB transmitter. The UWB receiver retrieves the data from the secondary node, the DSP processes the joint angle, and the UWB transmitter sends the calculated joint angle to the neural interface SoC, or external data logging system.

The neural interface SoC integrates: i) a 16-channel low-noise neural recording front-end, ii) a 16-channel programmable electrical stimulator, iii) a 10-bit successive approximation register (SAR) ADC, iv) a UWB wireless transceiver, v) a digital controller for generating timing and control signals, and vi) peripheral modules for power and analog references \cite{jan2016,liu2015PennBMBI}. The neural stimulator is designed to deliver biphasic charge-balanced stimulation in current-mode. The stimulation current amplitude is programmable. There are two modes designed for stimulating using clinical macroelectrodes and high-impedance microelectrodes. In the high-current mode designed for macroelectrodes, the stimulation current is programmable from 0 to 2mA; in the low-current mode designed for microelectrodes, the stimulation current is programmable from 0 to 200$\mu$A.

Local field potentials (LFPs) and action potentials (APs) are commonly used as control signals in sensorimotor BMI applications \cite{andersen2004}. In this work, the neural recording front-end has been designed with two modes to record LFPs and APs. The LFP mode has a bandwidth of 0.3Hz-1kHz, with a low noise floor; the AP mode has a bandwidth of 100Hz-6kHz, with a relaxed noise floor. The ADC has been designed with a sufficient dynamic range ($>$45dB) and a sampling rate ($>$10kSps/channel) to capture the amplified neural signals.

\section{Sensors and Circuits Design}
\subsection{Design of the Optical Tactile Force Sensor}
\begin{figure}[!ht]
  \centering
  \includegraphics[width=0.42\textwidth]{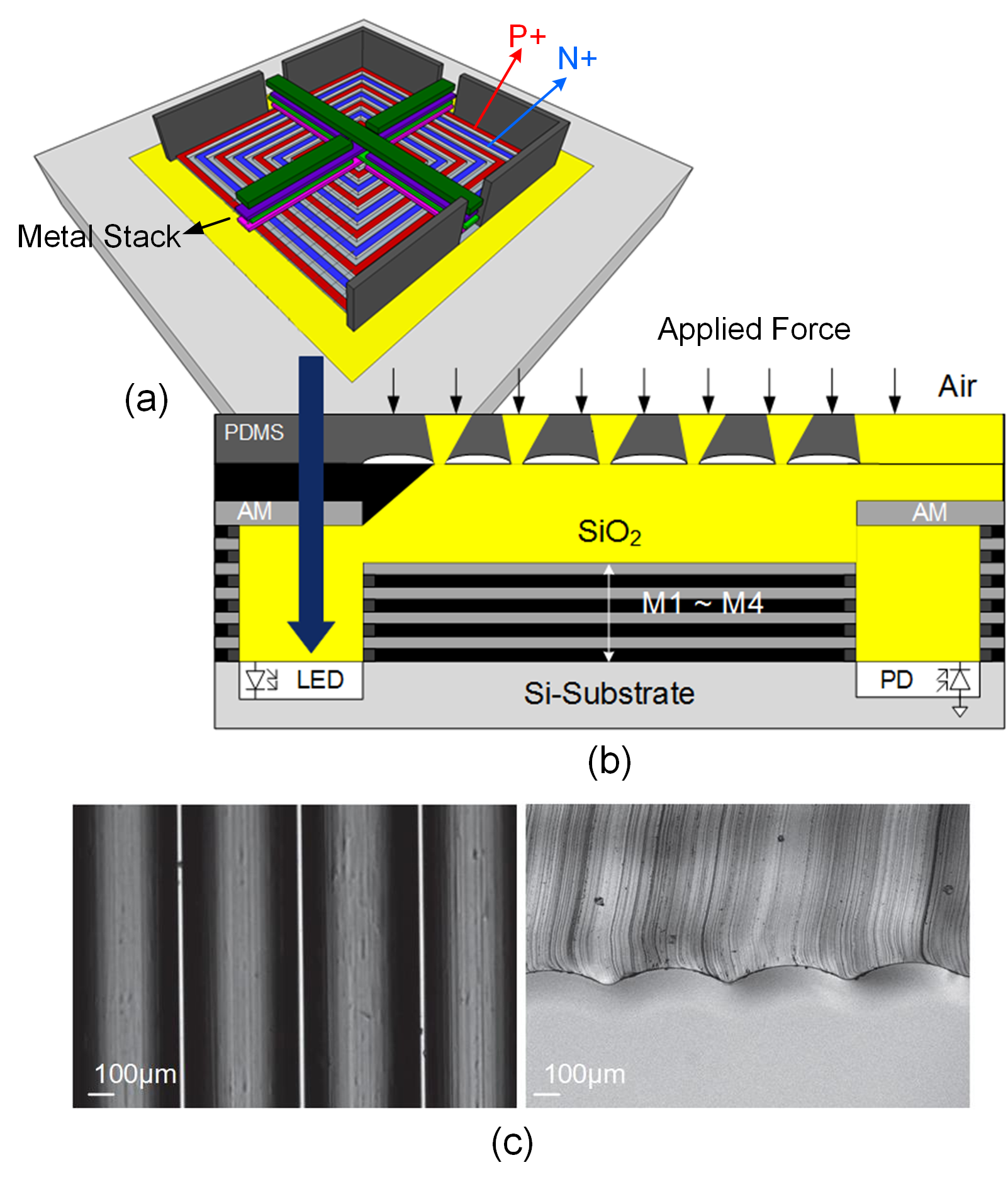}\\
  \caption{Illustration of optical force sensor. (a) 3-dimensional view of the silicon LED, which uses interdigitated P+ N+ rings inside an N-well. (b) Side view of the optical force sensor (not to scale). (c) Micrographs of the top (left) and side (right) views of the fabricated PDMS membrane with an inverse-lenticular structured surface.}\label{sensor_ckt}
\end{figure}

Fig. \ref{sensor_ckt} illustrates the optical force sensor design. A 600$\mu$m PDMS membrane is placed on top of the $SiO_2$ in standard CMOS die. The PDMS membrane is designed with an inverse-lenticular structured surface. As a result, the contact area between the PDMS and the $SiO_2$ is minimal when no force is applied, and the contact area increases with the applied force. During operation, a Si-LED emits light into the $SiO_2$ optical waveguide channel. A certain amount of light internally reflects and reaches the photodiode on the other side of the waveguide. The amount of the escaped light depends on the contact area of the PDMS and the $SiO_2$. Thus the readout of the photodiode changes monotonically with the applied force.

The PDMS material (SYLGARD 184, Dow Corning Co.) is composed of a 1:10 mixing ratio of the curing agent. The PDMS mixture is cast on a polystyrene lenticular lens board with a pitch of 20$\mu$m. After degassing for 30 minutes, the PDMS mixture and lenticular lenses molds are cured for 3 hours at 65$^{\circ}$C. Finally, the PDMS membrane is carefully peeled off the mold. The PDMS membrane is cut into 800$\mu$m by 800$\mu$m pieces and placed on top of the CMOS chip with the inverse-lenticular strips perpendicular to the direction of the optical waveguide.

The Si-LED is designed using interdigitated P+ N+ rings inside an N-well. The size of the Si-LED is 80$\mu$m by 80$\mu$m. On the other side of the optical waveguide, a photodiode is designed using P+ and N-well with an active area of 80$\mu$m by 80$\mu$m. The readout circuit uses a 3-transistor active pixel structure. The optical waveguide channel has a size of 200$\mu$m by 600$\mu$m. The sidewalls of the Si-LED, the photodiode and the waveguide channel are shielded by stacked metal layers and vias for minimum light leakage. The bottom side of the optical waveguide channel is elevated to four metal layers, which effectively prevent the light from being absorbed by the silicon substrate. It also reduces the path length of totally internally reflected light traveling inside the channel by reducing the thickness of the $SiO_2$ layer in the channel, which effectively reduces the light loss in the $SiO_2$ medium.

\subsection{Design of the Electrogoniometer}
A dual-accelerometer system is used for measuring the joint angle of two rigid body segments, $S_1$ and $S_2$. The mounting of the two accelerometers are illustrated in Fig. \ref{sys_intro}.

Define $\widehat{n}_g$ as a unit normal vector of gravity. The measured accelerometer vector is defined as $\vec{A_1}$, $\vec{A_2}$ $\in\mathbb{R}^3$, with respect to the local coordinate system. The cosine of the joint angle, $\theta$, can be written as:
\begin{equation}
cos\theta = \frac{\vec{A_1}\cdot\vec{A_2}}{||\vec{A_1}|| ||\vec{A_2}||}
\end{equation}\label{eq1}
Here we define $\Gamma_1$ and $\Gamma_2$ as
\begin{equation}
\Gamma_1 = \left [ \vec{A_1}\cdot\vec{A_2}\right ]^2
\end{equation}\label{eq2}
and
\begin{equation}
\Gamma_2 = \left[||\vec{A_1}|| ||\vec{A_2}||\right]^2
\end{equation}\label{eq3}

\begin{figure}[!ht]
\centering
  \includegraphics[width=0.32\textwidth]{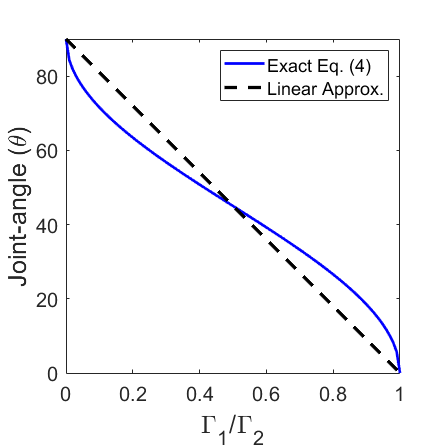}\\
\caption{Illustration of solving the joint-angle using the exact equation (Eq. 4) and the linear approximation (Eq. 5) implemented in this work.} \label{jtangle_approx}
\end{figure}

The exact joint-angle $\theta$ can be solved based on the Eq. (1-3) as
\begin{equation}
\theta = cos^{-1}(\sqrt{\frac{\Gamma_1}{\Gamma_2}})
\end{equation}\label{eq4}
However, the implementation of Eq. (4) significantly increases the complexity and power dissipation of the DSP hardware. In this work, we propose to use a simple linear equation to approximate the Eq. (4). The linear equation is given by
\begin{equation}
\theta' = \alpha (\frac{\Gamma_1}{\Gamma_2}) + \beta
\end{equation}
where $\alpha$ = $\beta$ = $\pi/2$.
The Eq. (4) and (5) are plotted in Fig. \ref{jtangle_approx} for comparison.

\begin{figure}[!ht]
\centering
  \includegraphics[width=0.5\textwidth]{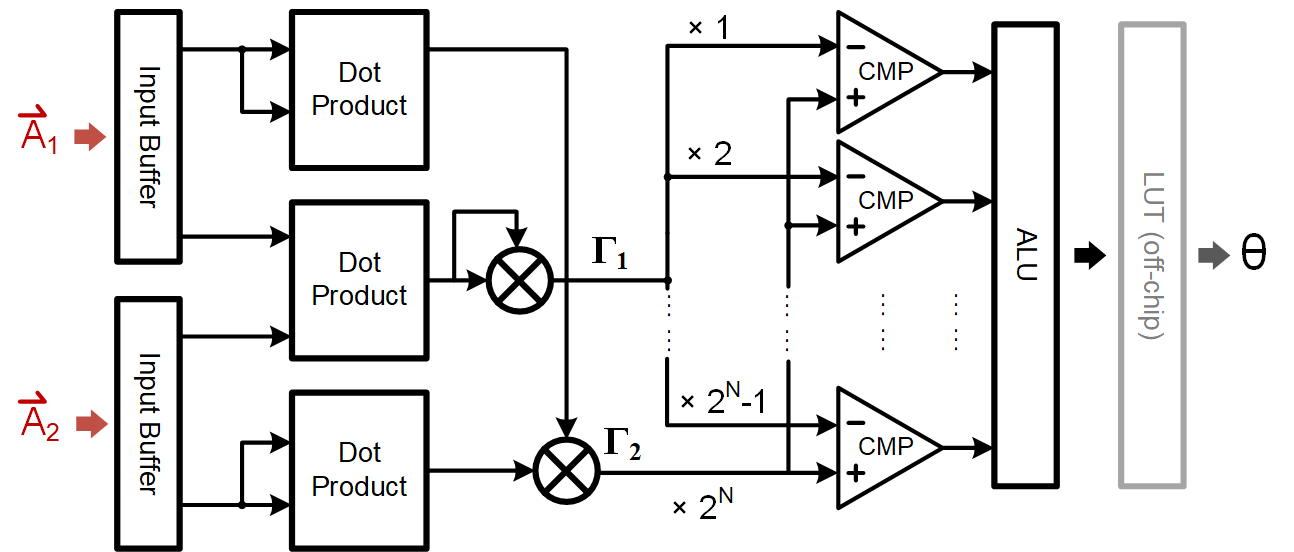}\\
\caption{Block diagram of the real-time joint angle calculation performed by the on-chip DSP. A linear N-bit approximation of $\theta$ is implemented on-chip, where N is 4. An off-chip look-up table (LUT) can be used to find the $\theta$ value.} \label{jtangle_arch}
\end{figure}

Fig. \ref{jtangle_arch} shows the block diagram of the hardware implementation.
During the operation, the two 6-bit digitized vectors $\vec{A_1}$ and $\vec{A_2}$ are first sent into the DSP for computing three dot products ($\vec{A_1}\cdot\vec{A_1}$), ($\vec{A_2}\cdot\vec{A_2}$), and ($\vec{A_1}\cdot\vec{A_2}$). The results are then used to produce $\Gamma_1$ and $\Gamma_2$.

To further reduce the computational cost, $\Gamma_1$/$\Gamma_2$ is quantized in 4-bit by a 16-level digital comparison instead of a full divider.
Specifically, $\Gamma_1$ is amplified in parallel from $\times$1 to $\times$15, and $\Gamma_2$ is amplified by $\times$16. The amplified versions of $\Gamma_1$ and $\Gamma_2$ are then compared in 15 individual digital comparators. At last, the comparators' output is fed into an arithmetic logic unit (ALU) for computing the linear equations Eq. (5). The final output is an estimated version of the joint-angle $\theta$.

Since the function of Eq. (4) is monotonic, the $\theta$ value can also be calibrated off-chip using a look-up table (LUT). The accuracy of the proposed implementation is mainly limited by the quantization level. This is an intentional design choice because a 5.625$^{\circ}$ joint-angle resolution is sufficient for the sensory encoding purpose in this work.

\subsection{Design of the Bidirectional Neural Interface}
\begin{figure}[!ht]
  \centering
  \includegraphics[width=0.45\textwidth]{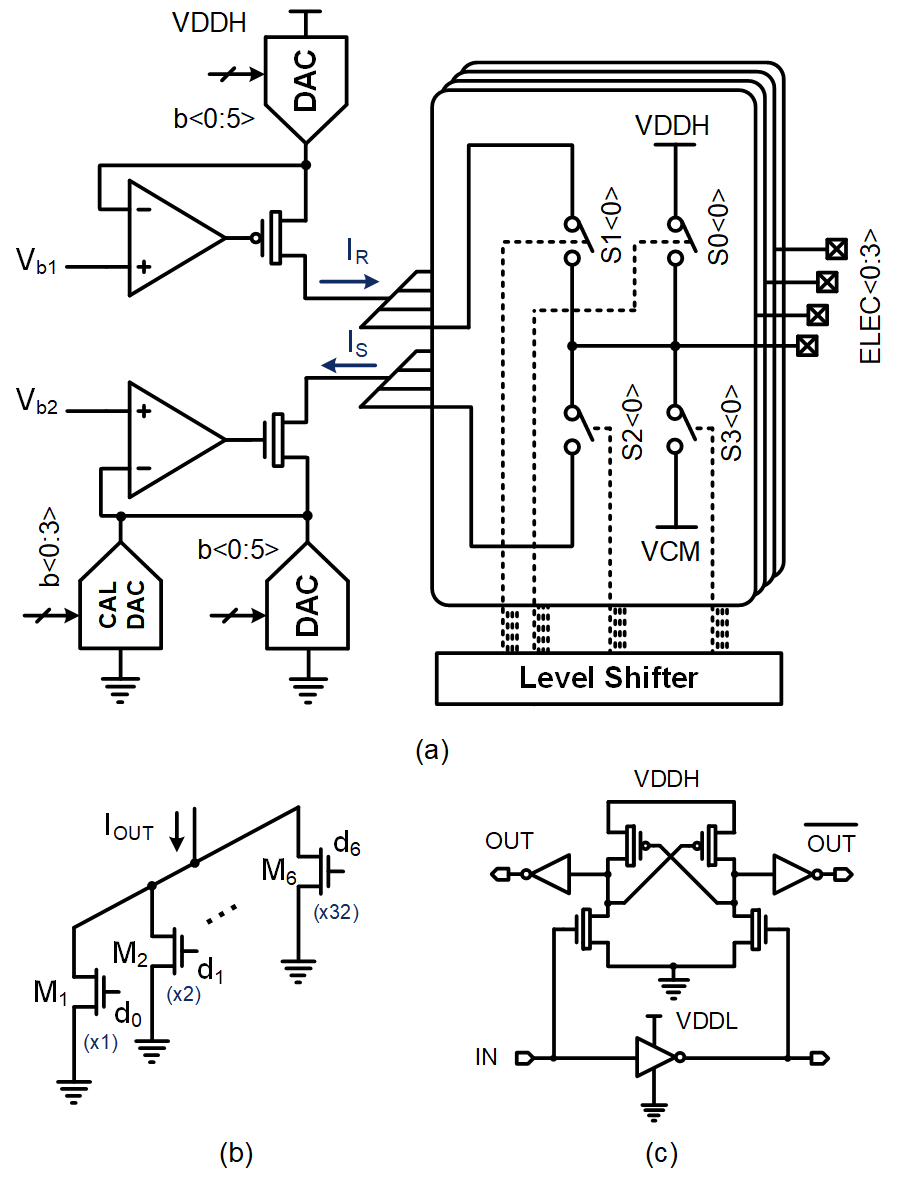}\\
  \caption{Circuit diagrams of the neural stimulator. (a) the output stage with a 6-bit DAC and a 4-bit calibration DAC, (b) the binary weighted current DAC, and (c) the level shifter.}\label{stim_ckt}
\end{figure}

The neural interface SoC integrates 16 independent channels for bidirectional neural stimulation and recording. Fig. \ref{stim_ckt} (a) shows the circuit schematic of the stimulation driving site, which is shared between a group of four channels. The timing of each driving site can be individually programmed. The stimulator can perform both monopolar stimulation and bipolar stimulation.
In the monopolar stimulation mode, one electrode is selected by a multiplexer; in the bipolar stimulation mode, two electrodes are selected from either the same or different driving sites \cite{liu2016pid}. Two 6-bit digital-to-analog converters (DACs) are designed to generate cathodic (sink) and anodic (source) stimulation currents. Each DAC consists of binary-weighted current sources, as shown in Fig. \ref{stim_ckt} (b). An additional 4-bit DAC is used for calibrating the static mismatches between the cathodic and anodic currents. The current of the sink DAC is intentionally reduced for a single-side calibration. A current comparator is integrated on-chip for calibration purposes. Regulating amplifiers are used to boost the output impedance.

The stimulator has two modes: in the high-current mode, the output current range is 0-2048$\mu$A with a programmable step current of 32$\mu$A; in the low-current mode, the output current range is 0-255$\mu$A with a programmable step current of 4$\mu$A. Thick oxide devices are used to tolerate high stimulation voltage. Fig. \ref{stim_ckt} (c) shows the level shifters used to convert a low-voltage digital signal to a high-voltage switch control signal. The whole stimulator module can be gated to minimize power leakage.

\begin{figure}[!ht]
  \centering
  \includegraphics[width=0.48\textwidth]{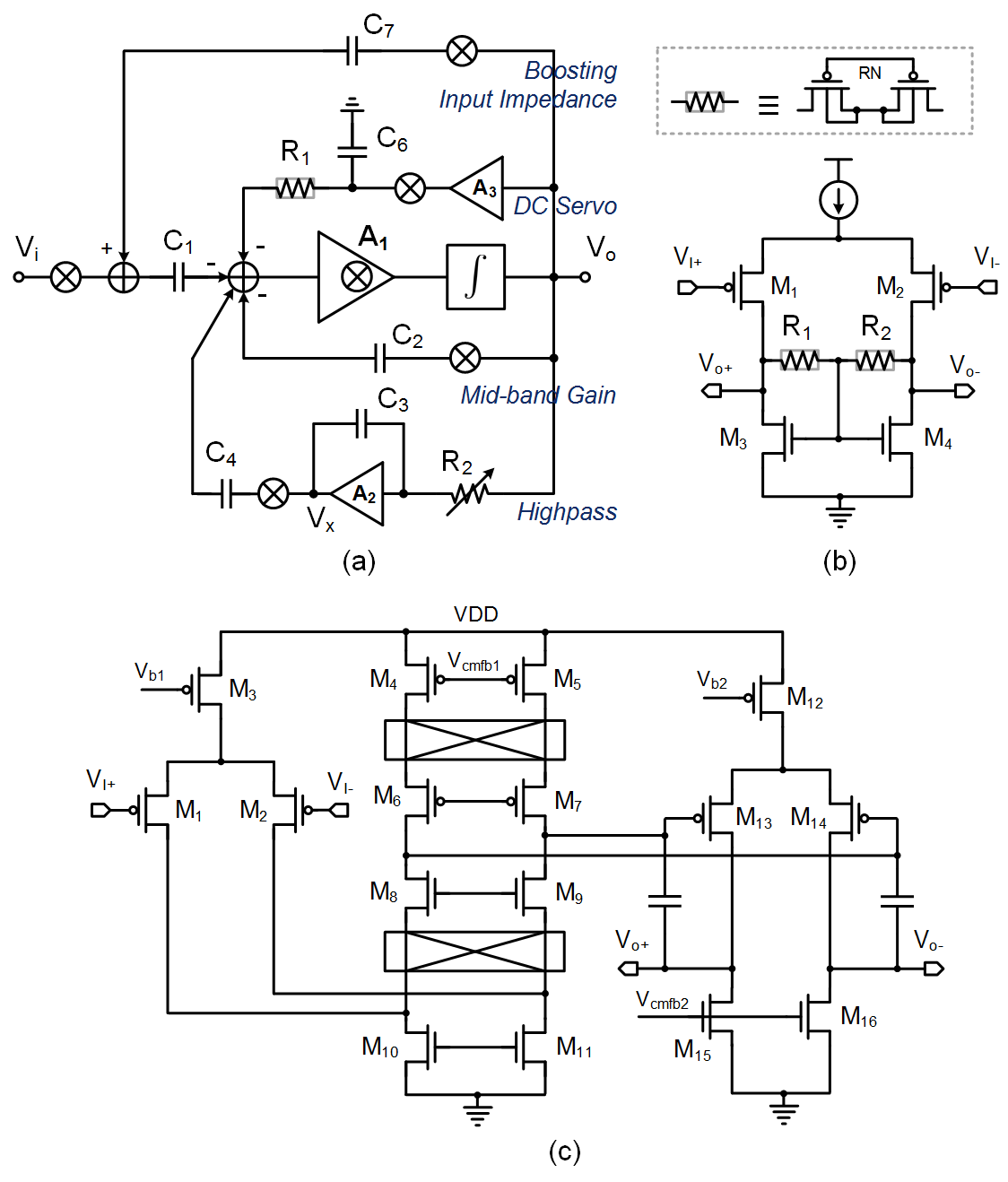}\\
  \caption{Circuits schematics of the neural recording front-end, including (a) the capacitively-coupled chopping amplifier. Single-ended structure is used for illustration, (b) the fully-differential amplifier used in the highpass and DC-servo loop, and (c) the two-stage low-noise transconductance amplifier with chopping switches.}\label{afe_ckt}
\end{figure}

The recording channel consists of a low-noise amplifier, a programmable transconductance-capacitance (Gm-C) bandpass filter, and a programmable gain amplifier (PGA). The key circuit diagrams of the low-noise neural recording front-end are shown in Fig. \ref{afe_ckt}. The low-noise amplifier uses capacitive feedback to set the gain. The core operational transconductance amplifier (OTA) A1 uses a folded-cascode topology. Chopping switches are integrated to remove the flicker noise \cite{liu2016csr,liu2016pid}. A capacitive positive feedback loop is used to boost the input impedance \cite{Fan2011}. A DC servo loop is used to remove the DC offset. Chopping is disabled when configured to record APs.

{\vspace{1cm}}

\subsection{Design of the Analog-to-Digital Converters}
Two ADCs have been designed in this work. First, a 10-bit SAR ADC is implemented in the neural interface SoC for the digitization of neural signals. Second, a 6-bit asynchronous LxADC is implemented in the sensor node for low-power sensor data digitization. SAR ADCs are suitable for low-power applications with moderate sampling rate requirements. In this work, a split capacitor array is used to reduce the area and power consumption. The capacitors are realized as a standard metal-insulator-metal (MIM) structure. A monotonic switching procedure is used to minimize the power consumption caused by unnecessary switching \cite{liu2016csr,liu2015biocas,liu2016pid}. 

\begin{figure}[!ht]
  \centering
  \includegraphics[width=0.45\textwidth]{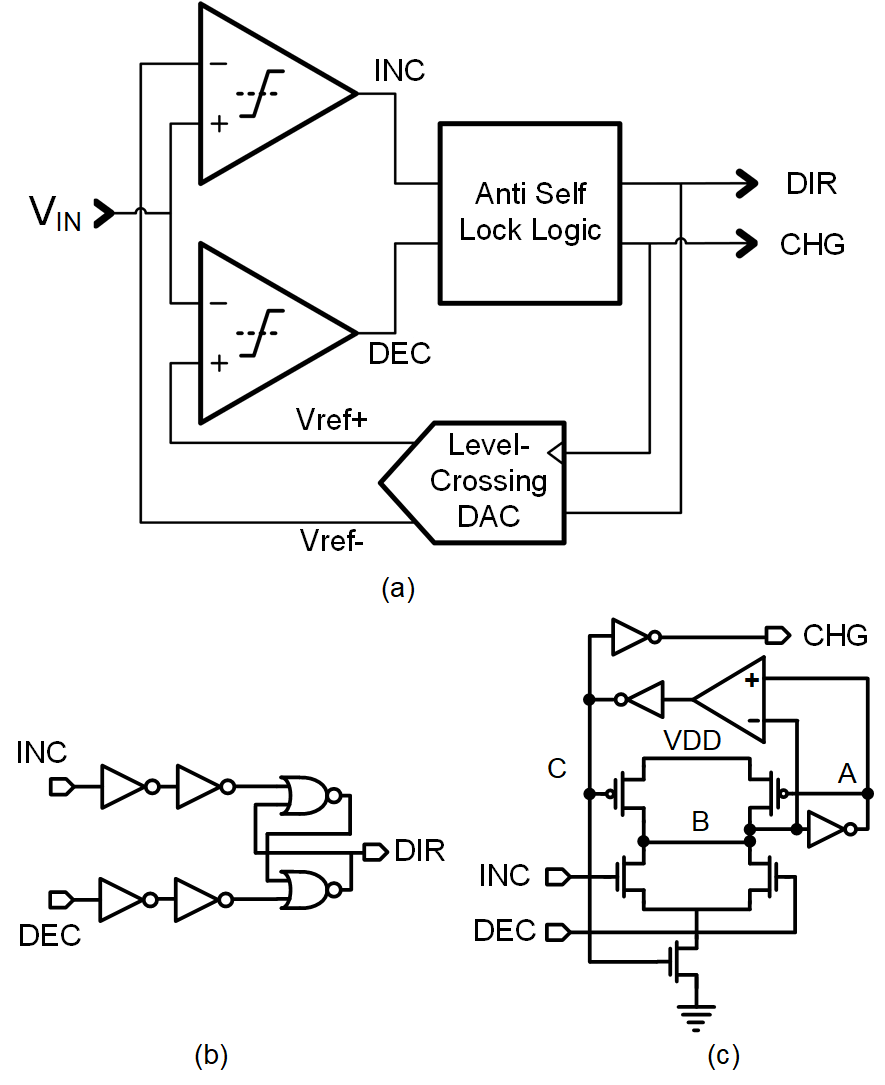}\\
  \caption{(a) Block diagram for the proposed LxADC. Circuit schematics for the (b) ``direction (DIR)" and (c) ``change (CHG)" signal generation modules for anti-self-locking operation.}\label{lxadc_ckt}
\end{figure}

Asynchronous continuous sampling ADCs have been introduced in ultra-low power applications in recent years \cite{tsividis2010event,tang2013continuous}. LxADCs have a low output data rate and signal-dependent power consumption \cite{tsividis2010event,schell2008A,weltin2013An}. A 200nW ultra-low power event-driven ADC with limited bandwidth and dynamic range has been developed \cite{zhang2013A}. An adaptive resolution asynchronous ADC has been proposed to improve data rate, at the cost of circuit complexity \cite{trakimas2011An}. Inspired by these designs, in this work, a low-power, high-speed asynchronous event-driven LxADC is developed for the sensor nodes. The block diagram of this LxADC is shown in Fig. \ref{lxadc_ckt} (a).

The LxADC tracks the changes of the input signal by comparing it with a set of hysteresis reference voltage levels using a pair of comparators \cite{zhu2015Design}. After reset, the LxADC first catches up with the input signal at its maximum speed. Once it catches up with the input signal, the continuous-time sampling mode will start. A 64-level reference voltage generator is used to provide an effective resolution of 6-bit. The comparator pair generates an ``increase (INC)'' or a ``decrease (DEC)'' signal when the input voltage crosses the level of the upper or the lower reference voltage, respectively.

\begin{figure}[!ht]
  \centering
  \includegraphics[width=0.4\textwidth]{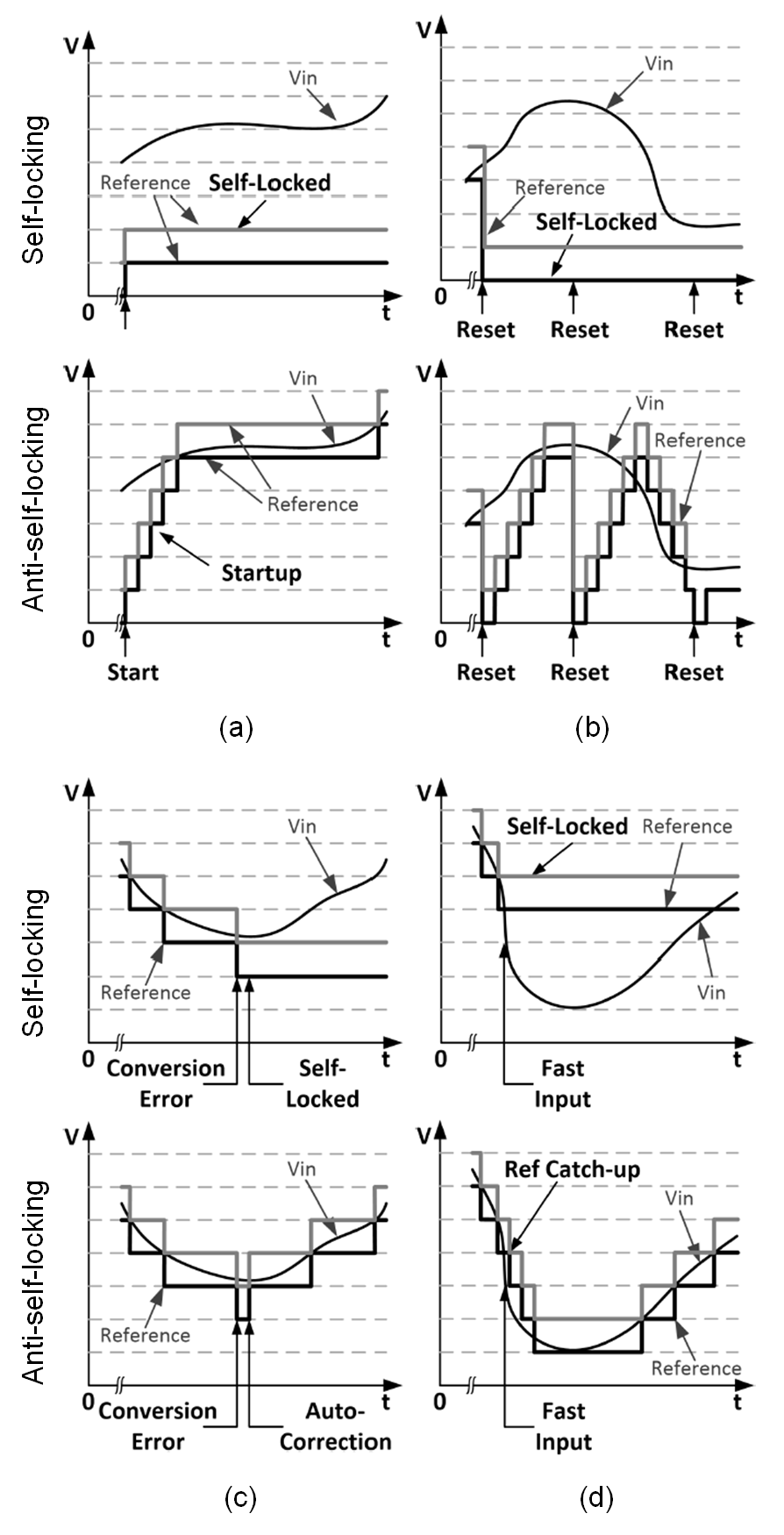}\\
  \caption{Illustration of self-locking conditions of the LxADC (upper rows) and anti-self-locking behavior of the proposed design (lower rows). Typical scenarios including operations during (a) ADC startup, (b) asynchronous resets, (c) conversion error, and (d) when the input signal changing faster than the LxADC's bandwidth.}
  \label{selflock}
\end{figure}

Self-locking is a critical issue in conventional LxADC designs. It occurs when the reference voltages, selected by the shift register, lose the capability for continuous tracking of the analog input voltage. Self-locking status often happens during 1) the circuit start-up phase, 2) circuit (shift register) reset phase, 3) conversion error, or 4) when the input signal changes faster than the circuit can respond. Fig. \ref{selflock} illustrates the scenarios of self-locking without and with the proposed anti-self-locking circuit. When the LxADC enters the self-locking state, either the ``INC'' or the ``DEC'' signal stays high without a rising edge generated to drive the shift register. A LxADC can only get out of the self-lock status if the input signal goes back in between the selected upper and lower reference voltages.

To release LxADCs from the self-locking state, anti-self-locking signals ``direction (DIR)" and ``change (CHG)'' are generated, as shown in Fig. \ref{lxadc_ckt} (b) and (c), respectively. The ``DIR" signal controls the direction of the shift register. In normal operations, the ``CHG" generation module passes the rising edges of the ``INC" or ``DEC" signal to the ``CHG" output. If the LxADC enters a self-locking state, the self-controlled delay loop formed by nodes B, A and C, generates rising edges of ``CHG" signals to drive the shift registers to update the reference voltage selection, until the LxADC exits. 
The maximum response speed of the LxADC is determined by the delay of the self-controlled loop in the regenerative ``CHG'' signal generation circuit. In order to avoid overshooting conversion of the LxADC, the delay of the self-controlled delay loop is designed to be longer than the conversion time of the input signal comparators.

The comparator, which is biased with a 30nA current, will delay the change of the voltages at nodes A and B at its input to the change of its output node by 0.5$\mu$s. A ``CHG'' signal is associated with the output of the comparator. The delayed change at node C will evoke the recharge of node B and discharge of node A, and hence the output of the comparator will be delayed again by 0.5$\mu$s to make node C high. If either the ``INC'' or the ``DEC'' signal is high at the moment when node C is recharged, node B will be discharged again and the loop will then generate another pulse for the ``CHG'' signal. With the regenerative ``CHG'' signal generation circuit, the proposed LxADC will be immune to self-locking states. Hence, the proposed LxADC features better robustness and makes an asynchronous reset function possible.

\subsection{Design of the Ultra-Wide Band Wireless Transceiver}
\begin{figure}[!ht]
\centering
  \includegraphics[width=0.45\textwidth]{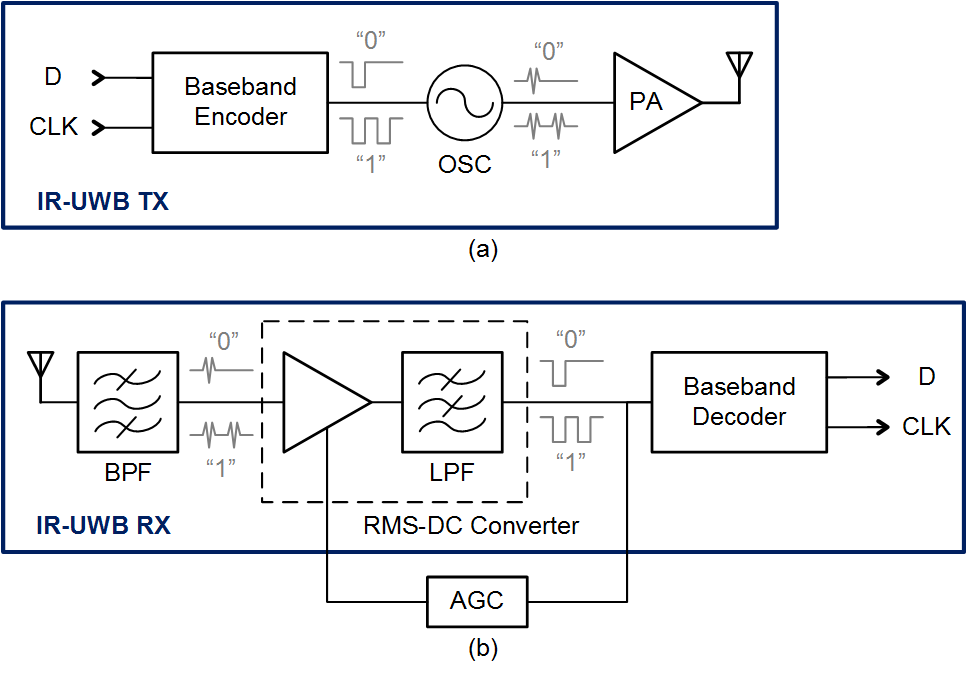}\\
\caption{Block diagram of the UWB (a) transmitter and (b) receiver.} \label{uwb_block}
\end{figure}

Impulse-radio UWB transceivers are widely used in near-range, power-sensitive applications. They are especially suitable for short-range wireless biomedical systems \cite{dokania2011low,zhang2018miniature,Benamrouche2017}, given the simple circuit structure, low power consumption, and high data rate. The block diagram of the UWB transceiver designed in this work is shown in Fig. \ref{uwb_block}.

The transmitter integrates a baseband generator, a RF pulse generator, and a power amplifier (PA). The baseband generator modulates digital input data into different numbers of short pulses. The pulse width is tunable under different data rates or transmission duty cycles. The RF pulse generator upconverts the short pulses to RF frequency. The oscillation frequency of the ring oscillator is tunable over a range of 100MHz. The RF pulse generator was implemented as a ring oscillator with a programmable number of stages.

In the receiver, the RF signal is first bandpass filtered at its corresponding operating frequency and then amplified by a low-noise amplifier. The output is fed into a RF power to root-mean-square voltage (RF-RMS) converter for downconversion. A comparator recovers the baseband short pulses, and a digital pattern recognition logic circuit demodulates the recovered signal to data and clock \cite{zhu2015Design}.

\section{Experimental Results}
\begin{figure}[!ht]
    \centering
    \includegraphics[width=0.43\textwidth]{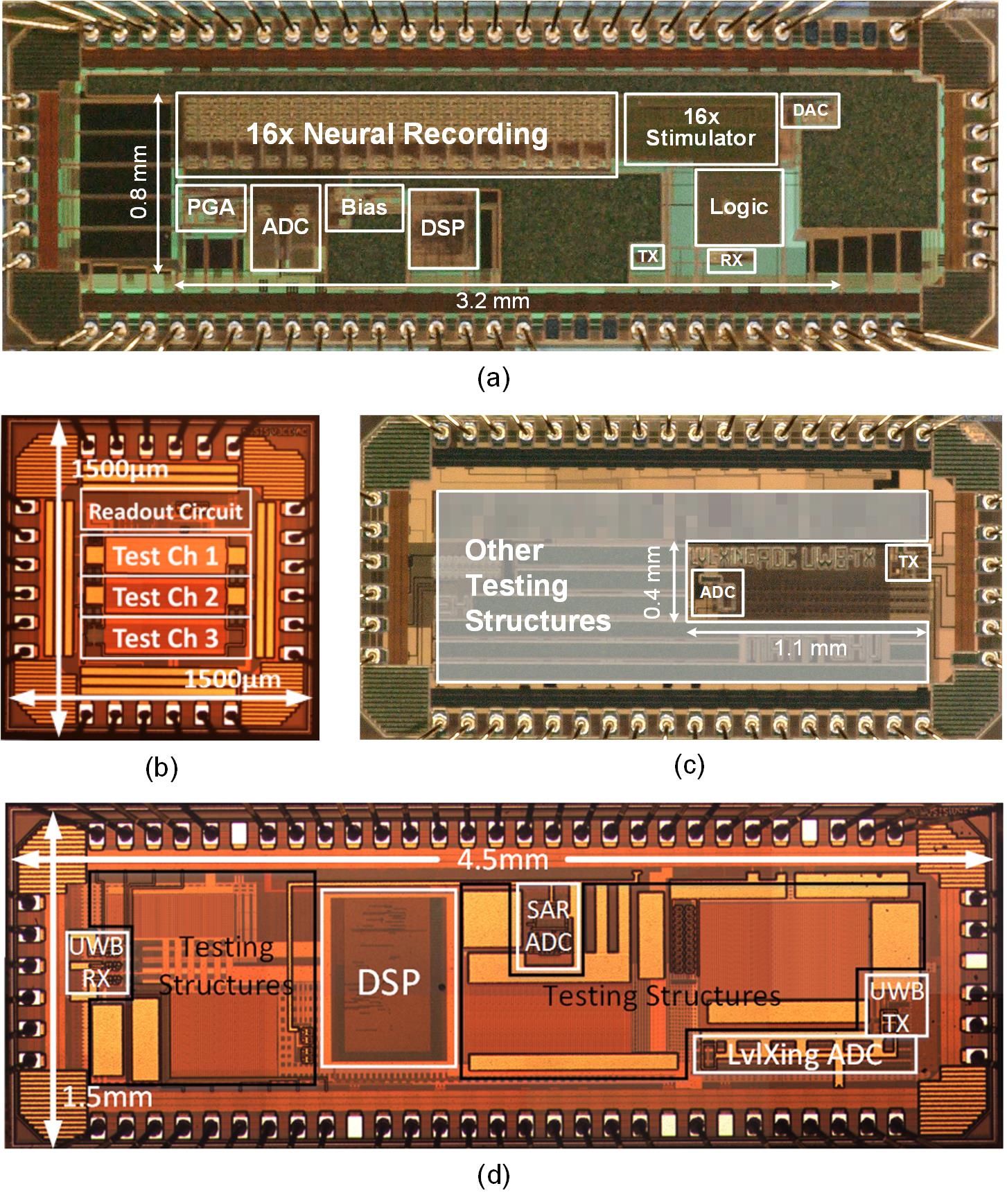}
    \caption{Micrographs of fabricated ICs: (a) the bi-directional neural interface SoC, (b) the optical force sensor, (c) the sensor interface node, and (d) the electrogoniometer. Major building blocks are highlighted.}
    \label{die_photo}
\end{figure}

The designed prototype ICs have been designed in Cadence Virtuoso and fabricated in 180nm CMOS technology. Microphotographs of the ICs are shown in Fig. \ref{die_photo}, with major building blocks highlighted. The optical force sensor occupies 1mm$\times$1mm, and the wireless sensor interface takes 1.1mm$\times$0.4mm. The size of the wireless electrogoniometer is 4mm$\times$1.1mm. The primary and secondary nodes are physically designed in the same silicon chip and specified through different configurations. The size of the neural interface SoC is 3.2mm$\times$0.8mm.

The fabricated ICs and off-the-shelf electronics were assembled on PCBs. Autodesk EAGLE was used for designing the PCBs. The off-the-shelf components in the devices mainly consist of a general-purpose microcontroller (MCU), a power management unit, and a lithium battery. The MCU model used in this work is Atmel ATxmega128A1U. The MCU is used for configuring the custom ICs, and is put in the power-down mode after the initial configuration for minimizing the power dissipation. A 3.7V 40mAh lithium-ion polymer battery from Adafruit is used for powering the device. The weight of the assembled BMI device and sensor nodes is less than 5g. In addition to these devices, a wireless computer interfacing dongle has also been designed for communication and data logging. A graphic user interface has been designed using MATLAB from MathWorks. The system design reuses aspects of our previous work \cite{liu2015PennBMBI,liu2016pid}.

\subsection{Bench Testing}
\begin{figure}[!ht]
  \centering
  \includegraphics[width=0.42\textwidth]{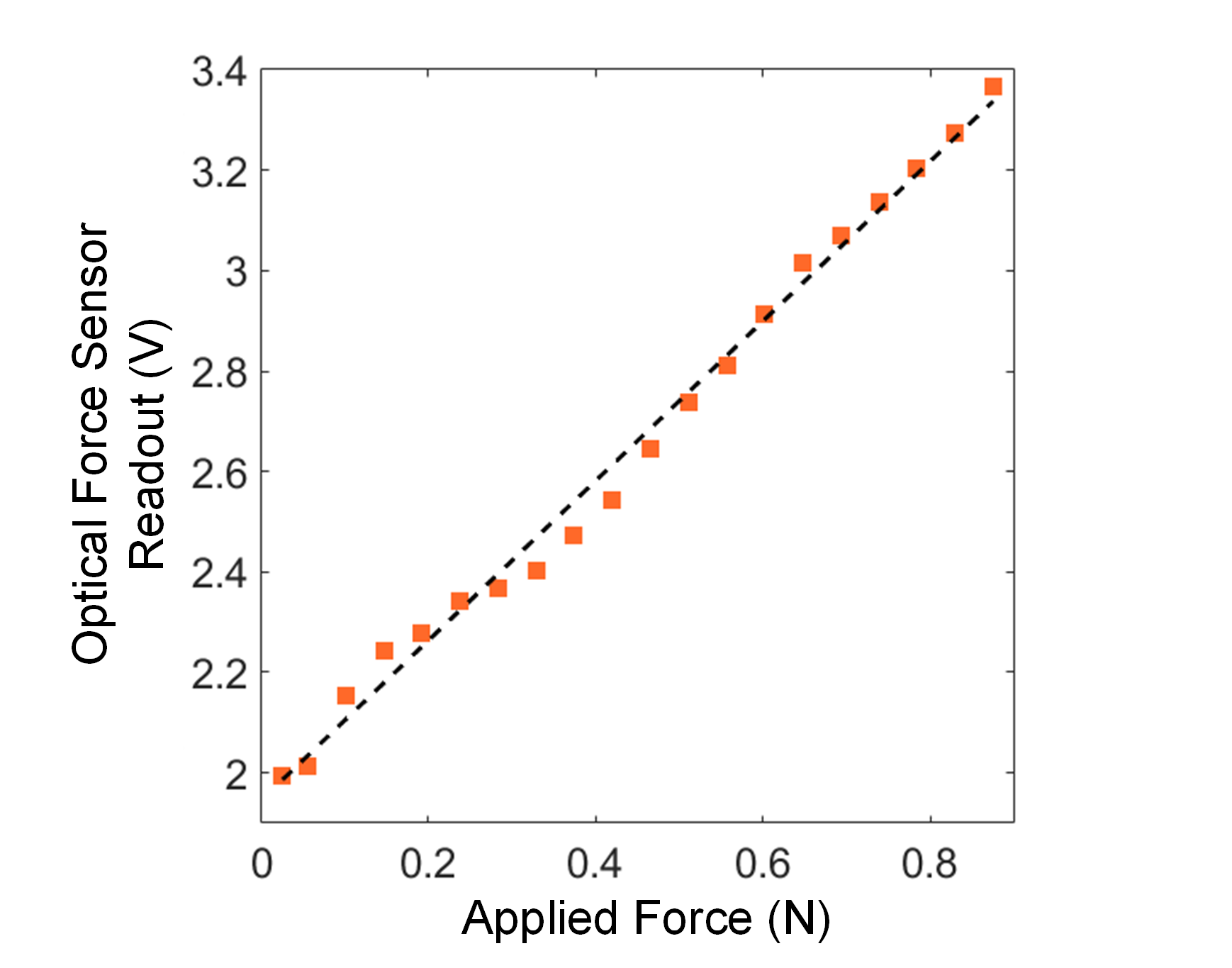}\\
  \caption{Measured outputs of the designed optical force sensor with respect to applied force without calibration. The measurement shows a good linearity with a $R^2$ value of 0.9892 within the force range of interest.}  \label{meas_sensor}
\end{figure}

\begin{figure}[!ht]
  \centering
  \includegraphics[width=0.34\textwidth]{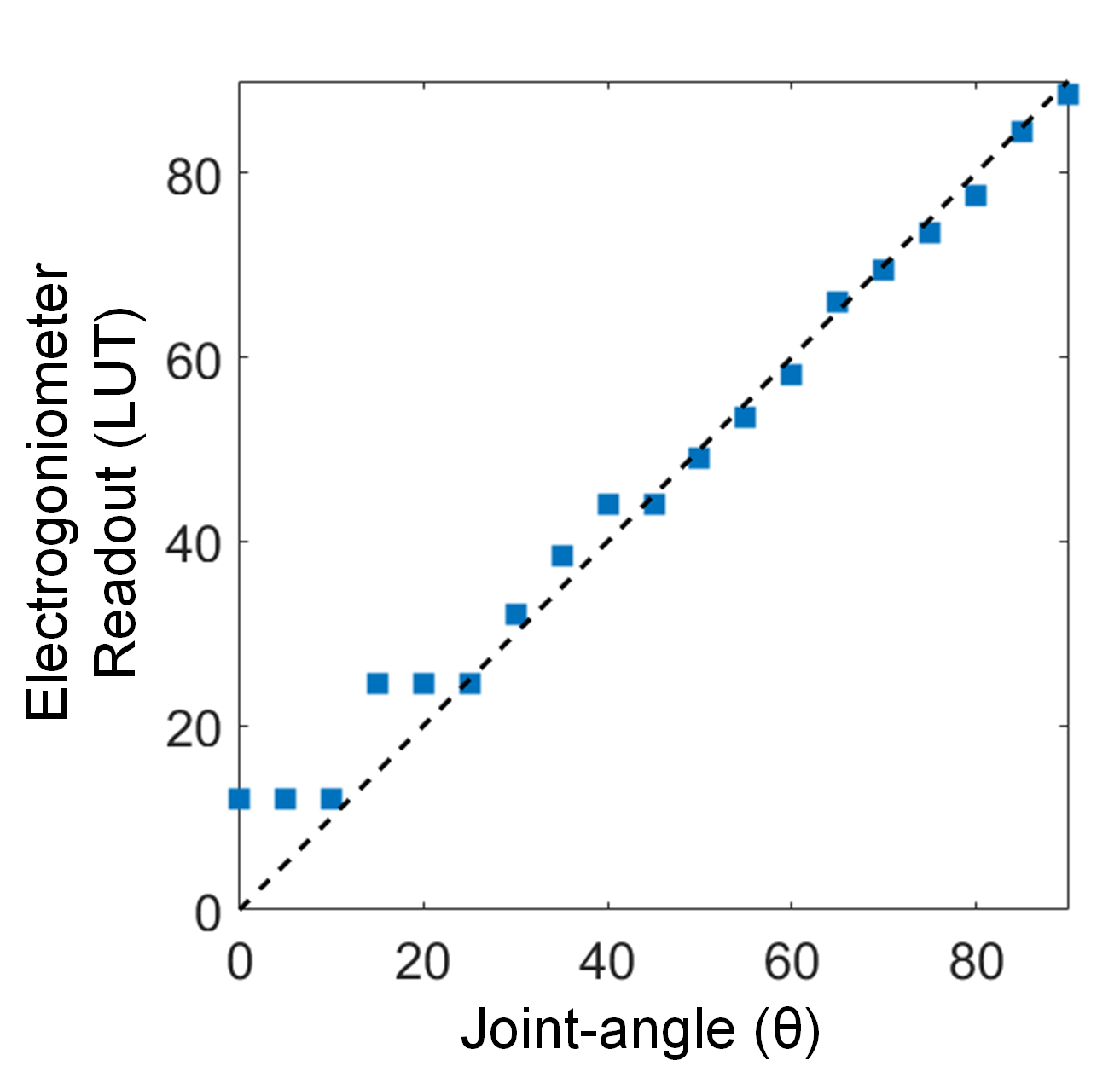}\\
  \caption{Measured outputs of the designed electrogoniometer versus different input angle ($\theta$). The output digital codes were corrected offline using a LUT.} \label{meas_jt}
\end{figure}

The fabricated silicon photodiode was characterized with and without the PDMS membrane. The experimental result suggests that the optical waveguide makes a negligible impact on the amount of light received by the photodiode. An external force ranging from 0 to 0.87N was applied to the optical force sensor, and the sensor's outputs were measured. Fig. \ref{meas_sensor} shows the results of the measurement before calibration.
The sensor response was monotonic with a nonlinearity of 2.53\%. A linear regression exhibits a $R^2$ value of 0.9892. A LUT can be employed to further improve the linearity by calibration. The sensitivity of the sensor is 13.6mN, corresponding to 2.125kPa with an 800$\mu$m$\times$800$\mu$m PDMS membrane area. Fig. \ref{meas_jt} shows the measured outputs of the electrogoniometer with different input angles. The nonlinear output codes were corrected offline by using a LUT. The readout of the electrogoniometer gives a sufficient resolution for the sensory restoration requirement in this work.

\begin{figure}[!ht]
    \centering
    \includegraphics[width=.46\textwidth]{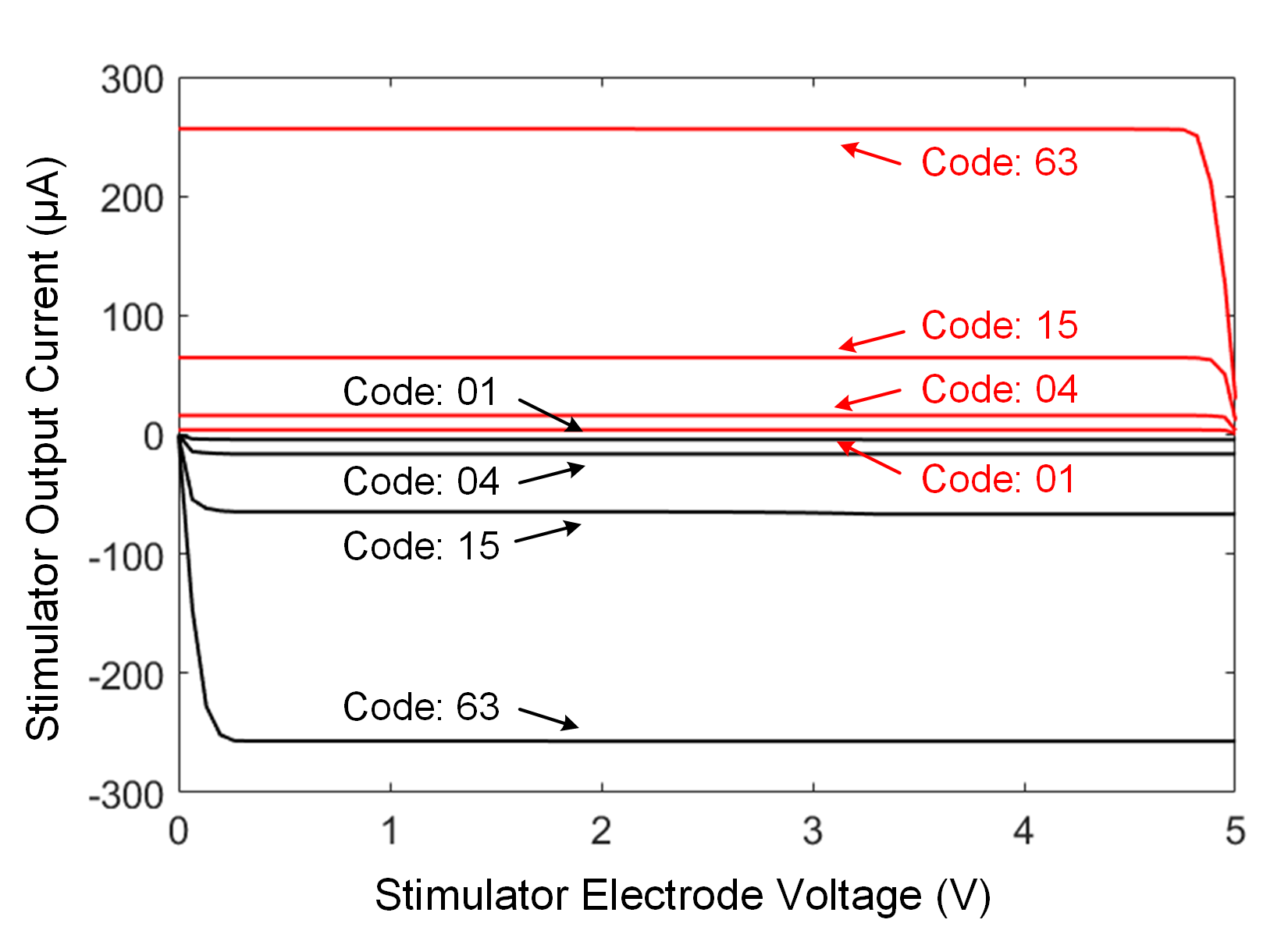}
    \caption{Measured stimulator output currents versus the output voltage of the electrode in different input codes. The compliance range of the designed stimulator was 95\% of the supply voltage.}
    \label{meas_stim}
\end{figure}

The designed neural stimulator has been tested for static and dynamic performance. The static mismatch between the cathodic and anodic current was 1.9\% before calibration, and 0.23\% after calibration. The stimulator's output currents were measured with DC output voltage. Fig. \ref{meas_stim} shows the measurement results in the low current mode with input codes 1, 4, 15, and 63. The compliance range of the stimulator output stage was 95\% of the supply voltage. Dynamic testing was conducted using a passive load consisting of a 1nF capacitor and 1k$\Omega$ in series. After calibration, a continuous 100-pulse train was delivered to the load without charging. The residue charges were measured, and the charge error during one stimulation pulse was calculated to the 0.35\% in this test.

\begin{figure}[!ht]
    \centering
    \includegraphics[width=.5\textwidth]{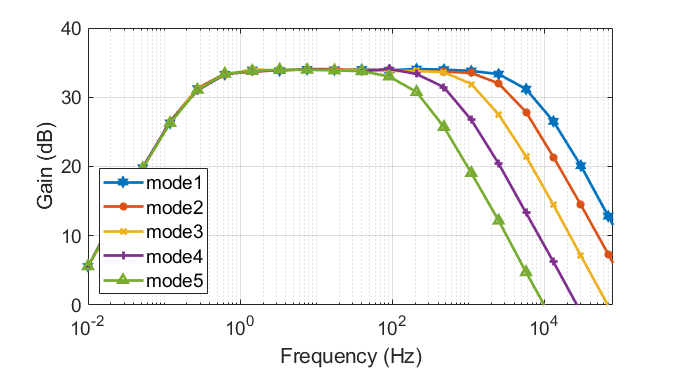}
    \caption{Measured frequency responses of the low-noise amplifier in different bandwidth modes. The bandwidth modes were configured by programming the Gm-C filter.}
    \label{meas_afe}
\end{figure}

The measured frequency responses of the low-noise amplifier in different bandwidth modes are shown in Fig. \ref{meas_afe}. The measured mid-band gain of the low-noise amplifier was 49.6 (34dB), and the common-mode rejection ratio (CMRR) was above 83dB in the worst case. The corner frequency of the low pass filter is programmable from about 200Hz to 6kHz. The measured input-referred noise of the low-noise amplifier in the 0.3Hz to 1kHz bandwidth is 1.58$\mu$V with chopping, which yields a noise efficiency factor (NEF) of 3.84 in LFP recording. The integral input-referred noise from 100Hz to 6kHz without chopping is 3.12$\mu$V, which yields a NEF of 2.82 in AP recording.

The LxADC works under a wide range of supply between 0.8V and 2.0V. The input voltage range is from 0.2V to VDD-0.15V. The measured output of the LxADC was synchronously recorded with a 1MHz sampling clock for storing, plotting, and post-digital signal processing. The maximum input signal slew rate of the ADC is 0.026V/us. The power consumption of the ADC is 5$\mu$W at a 0.8V supply with a 1kHz sinusoidal input signal. The SNDR of the LxADC is 46.2dB with a 5kHz sinusoidal input. The figure-of-merit (FoM) is calculated as 13pJ/conv. The SAR ADC was measured with a sampling rate of 1MSps. The peak DNL and INL are -0.49/+0.56LSB and -0.82/+0.77LSB, respectively. The spurious-free dynamic range (SFDR) is 76.54dB and the signal to noise and distortion ratio (SNDR) is 56dB. The effective number of bit (ENOB) is 9.01 bit. The FoM is calculated as 43pJ/conv.

The impulse radio UWB transmitter and receiver can operate over a supply range. The transmitter works from 1.2V to 2V, while the receiver works from 0.8V to 2V. The continuous RF output power is -33dBm, which can be increased to -13dBm with the high power PA on. The sampling clock frequency is tunable between 10MHz to 160MHz. The maximum data rate is 10Mbps, the measured power consumption of the transmitter is 4.6pJ/bit, and the receiver's power consumption is 1.12nJ/bit.

\subsection{In-vivo Testing}
\begin{figure}[!ht]
    \centering
    \includegraphics[width=.5\textwidth]{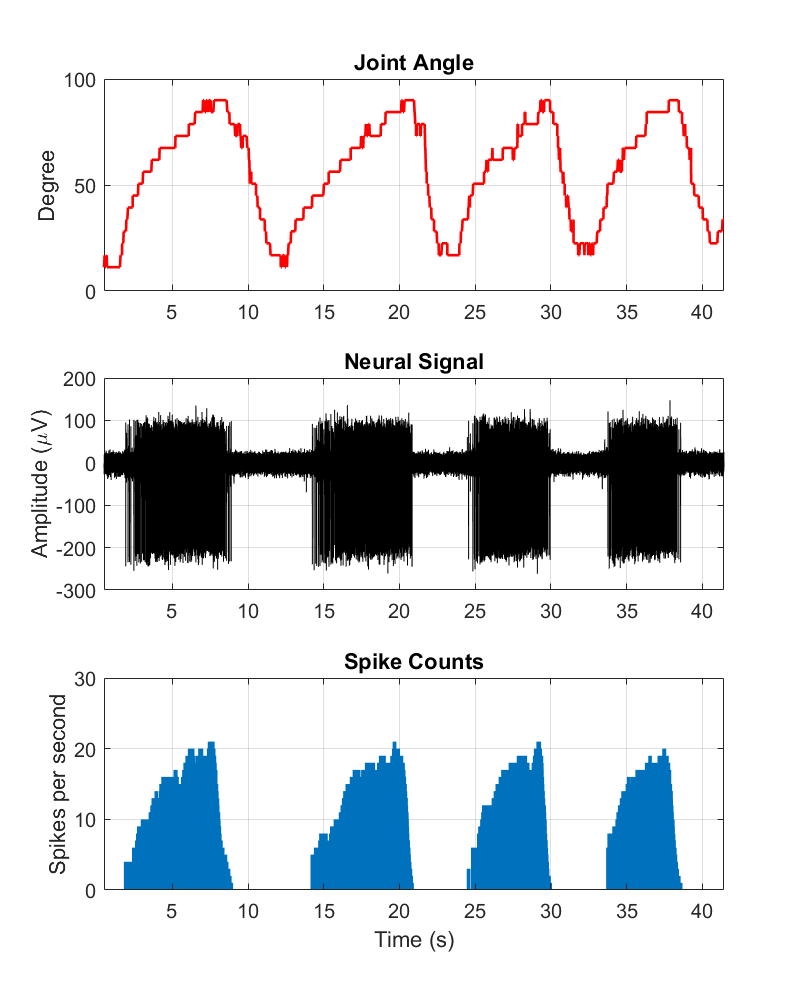}
    \caption{\emph{In-vivo} measurement of the electrogoniometer and the neural interface. (a) Measured joint angle, (b) recorded APs from a neuron, and (c) calculated spike rate of the APs using a moving window. A good correlation is shown between the joint-angle and AP firing rate. }
    \label{meas_joint}
\end{figure}

A subset of the functions of the SBMI system has also been tested \emph{in-vivo}. All tests were approved by the University of Pennsylvania Institutional Animal Care and Use Committee. A high-density electrode array was implanted in the somatosensory brainstem of a rhesus macaque \cite{Richardson2016Chronic}. Electrogoniometer nodes were fixed with elastic bands to the chest and upper arm of the sedated macaque to measure the shoulder angle. The neural interface was configured to record APs from a brainstem neuron sensitive to shoulder movements. A clear correlation between the joint angle and the AP firing rate was observed during shoulder abduction (Fig. \ref{meas_joint}).

\begin{figure}[!ht]
    \centering
    \includegraphics[width=.45\textwidth]{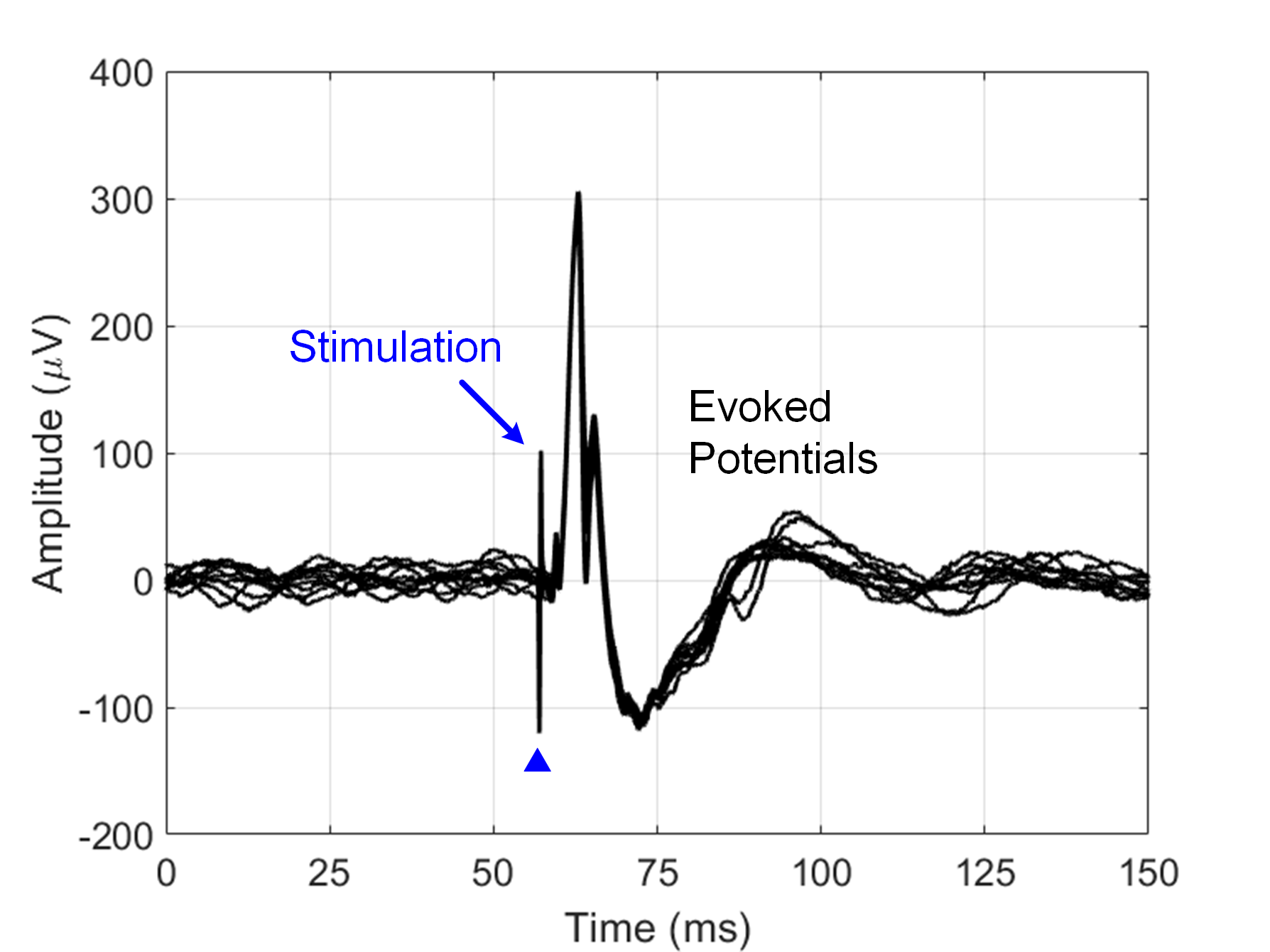}
    \caption{\emph{In-vivo} measurement of the bidirectional neural interface. Stimulation pulses were delivered to the somatosensory cortex of an anesthetized rat and the evoked potentials in the motor cortex were recorded. An overlay of 10 trials aligned with the stimulation time is shown in this figure.}
    \label{meas_spike}
\end{figure}

In addition, the bidirectional recording-stimulating capability of the neural interface was tested in an anesthetized Sprague-Dawley rat. A single stimulus pulse was repeatedly delivered to an electrode placed in the somatosensory cortex. The stimulus-evoked potential was recorded on a second electrode placed in the motor cortex. Fig. \ref{meas_spike} shows an overlay of the evoked potentials from 10 trials aligned with the simulation time. The experiment has demonstrated that the SBMI is capable of reliably evoking neurophysiological responses from stimulation of somatosensory areas.

During the sensation restoration operation, each wireless sensor node can individually trigger a pre-defined stimulus in the neural interface device. The stimulation parameters and activating electrodes are selected by experts and preloaded to the MCU. Either intensity or frequency of the stimulation can be linearly modulated by the sensor's output in real-time. The overall latency, including the wireless link, from the sensor node to the neural interface device was measured to be less than 2.5$\mu$s. In practice, the updating frequency of the stimulation parameters is typically limited by the sampling rate of the sensors.

To summarize the experimental results, key measured specifications of our SBMI system are listed in Table \ref{meas_table}. The power dissipation numbers listed in the table have been measured in the continuous modulation mode. The battery life of the devices is more than 24 hours. Finally, Table \ref{table_comparison} compares several key features of our system with recently reported BMI systems. Note that the noise performance in the LFP recording mode has been listed for comparison.

\begin{table}[!ht]
\centering
\caption{Measured Specifications Summary}
\label{meas_table}
\begin{tabular}{lll}
\hline\hline
Module & Specs & Performance \\ \hline
\multirow{7}{*}{\begin{tabular}[c]{@{}l@{}}Recording\\ Front-end\end{tabular}} & LNA Gain         & 34dB             \\
                                                                               & LNA Noise        & 1.58$\mu$V \\
                                                                               &                  & (0.3-1kHz, w/ chopping)\\
                                                                               &                  & 3.12$\mu$V \\
                                                                               &                  & (100-6kHz, w/o chopping)\\
                                                                               & CMRR             & \textgreater83dB \\
                                                                               & AFE Power            & 8.3$\mu$W per ch      \\ \hline
\multirow{6}{*}{\begin{tabular}[c]{@{}l@{}}Neural\\   Stimulator\end{tabular}} & Stim. Current    & 0 - 255$\mu$A/2mA          \\
                                                                               & Amplitude Res.   & 6-bit            \\
                                                                               & Pulse width      & 1$\mu$s - 250$\mu$s      \\
                                                                               & Stim. Frequency  & 0.5Hz - 300Hz    \\
                                                                               & Charge Error     & 0.35\%             \\\hline
\multirow{6}{*}{\begin{tabular}[c]{@{}l@{}}UWB \\ Tranceiver\end{tabular}}     & TX Output power  & -33dBm           \\
                                                                               & TX/RX min supply & 1.2V/0.8V        \\
                                                                               & Frequency        & 1.6 - 1.7GHz       \\
                                                                               & Max data rate    & 10Mbps           \\
                                                                               & UWB TX power     & 4.6pJ/bit        \\
                                                                               & UWB RX power     & 0.32nJ/bit       \\ \hline
\multirow{4}{*}{\begin{tabular}[c]{@{}l@{}}SAR/ \\ LxADC\end{tabular}}         & Sampling Rate    & 1M/5kHz          \\
                                                                               & ADC ENOB         & 9.0/6            \\
                                                                               & Power supply     & 1.8/0.8V         \\
                                                                               & LxADC power      & 13pJ/conv        \\ \hline
\multirow{2}{*}{\begin{tabular}[c]{@{}l@{}}Force\\ Sensor\end{tabular}}        & Sensitivity      & 13.6mN           \\
                                                                               & Non-linearity    & 2.53\% \\ \hline
\multirow{3}{*}{\begin{tabular}[c]{@{}l@{}}Power\\ Consumption\end{tabular}}   & BMI device       & 1.4mW            \\
                                                                               & Force Sensor     & 0.7mW           \\
                                                                               & Electrogoniometer      & 1.1mW (primary node) \\
                                                                               &     & 0.2mW (secondary node) \\ \hline\hline
\end{tabular}
\end{table}

\begin{table*}[!ht]
\centering
\caption{Comparison with Bidirectional Brain Machine Interface Designs}
\label{table_comparison}
\begin{tabular}{c|c|c|c|c|c|c|c}
\hline\hline
Reference &  \cite{Rhew2014}   &  \cite{cong2014} &  \cite{liu2015PennBMBI}      &  \cite{biederman2015}  &  \cite{liu2016pid} &  \cite{greenwald2016}     & This work \\ \hline
Publication  & 2014 JSSC & 2014 ESSCIRC & 2015 TBioCAS  & 2015 JSSC & 2016 TBioCAS & 2016 TBioCAS   & -   \\ \hline
Technology  & 180nm  & 0.25$\mu$m/90nm   & PCB    & 65nm  & 180nm   & 180nm   & 180nm  \\ \hline
AFE ch \#  & 4ch   & 32ch  & 3ch  & 64ch & 16ch & 4ch & 16 ch \\ \hline
AFE Noise   & 6.3$\mu$Vrms  & 100nV/rtHz  & 4.72$\mu$Vrms  & 7.5$\mu$Vrms  & 4.57$\mu$Vrms  & 1.0$\mu$Vrms  & 1.58$\mu$Vrms   \\ \hline
Bandwidth (Hz)  & 0.64-6kHz &100Hz & 0.05-6kHz & 10/1kHz-3k/8kHz & 0.3-7kHz &  0.25-250 Hz & \begin{tabular}[c]{@{}c@{}}0.3-1kHz \\ (or 100-6kHz)\end{tabular} \\ \hline
AFE NEF & 3.76 & Not reported & Not reported & 3.6 & 4.77 & 2.5 & 3.84 \\ \hline
Stim ch \# & 8ch Monopolar  & 16ch Monopolar & \begin{tabular}[c]{@{}c@{}}8ch Mono/\\     bipolar\end{tabular}   & 8ch Bipolar    & \begin{tabular}[c]{@{}c@{}}16ch Mono\\     /bipolar\end{tabular}       & 4ch                                                               & \begin{tabular}[c]{@{}c@{}}16ch Mono\\     /bipolar\end{tabular}     \\ \hline
\begin{tabular}[c]{@{}c@{}}Stim Supply (V)\end{tabular} & 5V & Not reported & +/-12V & Not reported & Not reported & 5V & 5.5V \\ \hline
\begin{tabular}[c]{@{}c@{}}Max. Output (I)\end{tabular} & 4410$\mu$A   & 12mA    & 10mA & 900$\mu$A   & 4mA   & 250$\mu$A  & 2mA  \\ \hline
Sensors  & -  & -   & \begin{tabular}[c]{@{}c@{}}Pressure\\     Accelerometer\\     Temperature\end{tabular} & -   & -   & -   & \begin{tabular}[c]{@{}c@{}}Pressure\\     Accelerometer\end{tabular} \\ \hline
ADC Mode          &  \begin{tabular}[c]{@{}c@{}}Pipeline\\   log-ADC\end{tabular}       & $\Sigma$$\Delta$ ADC & 12-bit SAR & 10-bit SAR  & \begin{tabular}[c]{@{}c@{}}SAR\\     Current-mode\end{tabular} & $\Sigma$$\Delta$ ADC & \begin{tabular}[c]{@{}c@{}}SAR/\\  LxADC\end{tabular} \\ \hline
ADC ENOB & 5.6 & 12 & Not reported & 8.2 & 9.1/7.9 & 9.4 & 9.0/6 \\ \hline
ADC FoM & Not reported & Not reported & Not reported & Not reported &  \begin{tabular}[c]{@{}c@{}}34.2fJ/conv-step\\ 10.7fJ/conv-step\end{tabular} & 7.6 pJ/conv & \begin{tabular}[c]{@{}c@{}}43pJ/conv\\ 13pJ/conv\end{tabular}\\ \hline
On-chip proc  & Custom DSP   & Custom CPU    & -     & Custom DSP    & Analog parallel   & -     & Custom DSP   \\ \hline
\begin{tabular}[c]{@{}c@{}}Wireless\\ Link\end{tabular}    & \begin{tabular}[c]{@{}c@{}}Custom \\ backscattering\end{tabular}      & -                                                             & \begin{tabular}[c]{@{}c@{}}Commercial\\     GFSK\end{tabular}                          & -      & \begin{tabular}[c]{@{}c@{}}Commercial \\ Bluetooth\end{tabular}        & -                                                                 & \begin{tabular}[c]{@{}c@{}}Custom \\ UWB\end{tabular}                \\ \hline
Closed-loop                                                & Yes                                                                   & Yes                                                           & Yes                                                                                    & Yes                                                             & Yes                                                                    & -                                                                 & Yes                                                                  \\ \hline
Application                                                & \begin{tabular}[c]{@{}c@{}}Deep brain \\     stimulation\end{tabular} & Generalized                                                   & Generalized                                                                            & \begin{tabular}[c]{@{}c@{}}Neuro-\\     modulation\end{tabular} & Generalized                                                            & Generalized                                                       & \begin{tabular}[c]{@{}c@{}}Sensory \\     restoration\end{tabular}   \\ \hline\hline
\end{tabular}
\end{table*}

 \vspace{1cm}

\section{Conclusion}
In this paper, a fully integrated wireless SBMI system has been presented. The system consists of a wireless bidirectional neural interface and custom-designed sensor nodes for transducing key somatosensory stimuli. A novel optical force sensor in standard CMOS with low-cost post-fabrication has been developed. Since the sensor is compatible with CMOS circuits, the sensor and all processing circuits can be integrated into a single chip in the future. The miniature design is compatible with tactile sensing on hands that have lost sensation due to injury. Key future issues to address are potential wireless sensor powering strategies and robustness of the design to repeated mechanical loading during daily use.

An electrogoniometer has been designed with custom circuits and low-cost accelerometers, which significantly reduces the power consumption compared with the strain gauges widely used in biomechanical studies. A custom-designed on-chip joint angle digital processor has been designed. The custom DSP minimizes the delay in joint angle calculation, which is critical in real-time sensory encoding paradigms. Again, future work could focus on the wireless power of these sensor nodes for a fully wireless SBMI. An asynchronous event-driven LxADC has been designed, which reduced the wireless data rate significantly compared with a conventional synchronous Nyquist-rate sampling system. The custom-designed impulse radio UWB wireless link achieves low power consumption in a small silicon area, which is especially suitable for short-range biomedical communication.

A bidirectional neural interface has been designed for neural stimulation and recording. The neural stimulator delivers biphasic charge-balanced stimulation with programmable current amplitude and frequency. The neural recorder amplifies LFP or AP signals with programmable gain and bandwidth. All of the ICs have been fabricated and evaluated in bench tests and \emph{in vivo}. Compared with state-of-the-art designs summarized in Table \ref{table_comparison}, this work demonstrates a novel integrated wireless system for sensation restoration, as well as novel circuit and sensor implementations. Based on the preliminary results, the proposed SBMI system provides a promising platform with which to test sensory encoding strategies in freely-behaving animal models and, in turn, advance next-generation, closed-loop neural prosthetics for individuals with paralysis.


\end{document}